# Spectral self-adaptive absorber/emitter for harvesting energy from the sun and outer space

Xianze Ao[1]*, Bowen Li[2]*, Bin Zhao[1], Mingke Hu[1], Hui Ren[2], Honglun Yang[1], Jie Liu[1], Jingyu Cao[1], Junsheng Feng[1], Yuanjun Yang[2], Zeming Qi[2], Liangbin Li[2], Gang Pei[1]† , Chongwen Zou[2]†

**Affiliations:**

[1]Department of Thermal Science and Energy Engineering, University of Science and Technology of China, Hefei, Anhui 230027, China.

[2]National Synchrotron Radiation Laboratory, University of Science and Technology of China, Hefei, Anhui 230029, China.

*These authors contributed equally to this paper.

†Corresponding author. Email: peigang@ustc.edu.cn; czou@ustc.edu.cn

**Abstract:** The sun (~6,000 K) and outer space (~3K) are the two important energy resources for human beings on Earth. The energy applications by absorbing solar irradiation and harvesting the coldness of outer space have received tremendous interest for energy utilization. However, combining these two functions in a static device for continuous energy harvesting is always unachievable due to the intrinsic infrared spectral confliction. Here, we realize a spectrally self-adaptive absorber/emitter (SSA/E) for daytime photothermal and nighttime sky radiative cooling modes depending on the phase transition of coated $VO_2$ layer at ~68 ℃. The 24-hour day-night test shows that the fabricated SSA/E has continuous energy harvesting ability as well as the improved overall energy utilization performance, showing great promising for future energy applications.

Heating and cooling are two significant end-use of energy in both industrial sectors and daily life. In the US, the demand of thermal energy at intermediate temperatures (120−220 °C) is larger than the total annual electricity generation from all renewable sources in 2016 (*1*). Besides, heating and cooling of buildings account for nearly 48% of the total energy used by buildings, making it the largest individual energy expense (*2*). Therefore, heat and coldness generated from renewable energy will have a considerable potential for global energy conservation and emission reduction. It is well-known that photothermal effect has been widely used to convert the solar energy to heat for various applications (*1*, *3*, *4*). Recently, energy harvesting from the cold outer space of 3 K by sky radiative cooling technology (*4-26*) has also been one of the most attractive research topics. This technology utilizes the cold outer space as a heat sink and terrestrial objects dissipate heat mainly through the 8−13 μm atmospheric window.

Considering the fact that there is a large demand for both heat and coldness of the society, combining these two functions of harvesting heat in the daytime and coldness in the nighttime using one device is highly desirable for the round-the-clock application. Though the physical principles of daytime photothermal (DPT) and nighttime sky radiative cooling (NRC) are similar, there exists intrinsic spectrally selective confliction. Specifically, the DPT absorber needs to have low emissivity in the mid-infrared wavelength band (2.5−25 μm), while NRC emitter needs to exhibit strong thermal emission within the atmospheric window (8−13 μm). Because of the incompatible mid-infrared selective confliction between two working modes, conventional DPT absorber and NRC emitter can't be simply integrated together for 24-hour continuous energy harvesting. In order to efficiently harvest heat in the daytime and coldness in the nighttime with an entirely passive solution, some research groups try to integrate the NRC and DPT into one device. While almost all of the studies focus on static-spectrum NRC-DPT structures (*27, 28*)，which has invariable spectral characteristic. Normally an ideal static-spectrum of this kind of NRC-DPT structure should have unity emissivity within solar radiation band and atmospheric window, while keeping zero value in other infrared bands (fig. S1). However, this idea structure has intrinsic defect.  Specifically, high emissivity within the atmospheric window can be beneficial to NRC, but it will sacrifice thermal performance of DPT because of the increased infrared radiation loss, showing the obvious and incompatible infrared spectrum conflict.

In the current study, we propose an integration device with switchable DPT and NRC functions to overcome the infrared spectrum conflict. According to the theoretical simulation, we fabricate a spectral self-adaptive coating (SSAC) based on vanadium dioxide ($VO_2$) phase transition material (*29-32*), whose spectral characteristics are automatically changed by its temperature. Experimental results show that this spectral-self-adaptive coating can be heated up to ~170 °C above the ambient temperature under solar intensities of ~800 W m$^{-2}$ in the DPT mode and cooled down to ~20 °C below the ambient temperature in the NRC mode. The excellent performance of the SSAC makes it promising to efficiently harvest heat and coldness from the universe all the time.

The principle of the spectral self-adaptive NRC-DPT hybrid utilization is presented in Fig. 1, A and B.  An idea NRC-DPT device should be an excellent radiative cooling emitter in nighttime and ideal photothermal absorber in daytime. It mainly aims to make the device cooled at nighttime by keeping a high emissivity in the atmospheric window (8−13 μm), while making the device heated at daytime by fully absorbing incident solar radiation and avoiding the infrared radiation loss. Thus the infrared spectroscopy, especially within the atmospheric window, should be adjustable to improve the energy harvest efficiency. According to this spectra selectivity rule, the proposed SSAC can be achieved by a thermochromic $VO_2$ layer as shown in Fig. 1, C and D. This dynamical spectra switching is controlled by the device temperature, which is originated from the phase transition of $VO_2$ layer at the critical temperature of ~68 °C. Specifically, during the nighttime, the SSAC device has a lower temperature, thus the $VO_2$ layer keeps the insulator state and the SSAC acts as a good radiative cooler (Fig. 1A). While going to the daytime, the temperature of the SSAC rises after absorbing solar radiation. If the temperature continuously goes higher than the critical temperature of $VO_2$ film, the $VO_2$ layer will convert from an insulating phase to a metallic phase. As a result, the SSAC switches off the radiative cooling and becomes a solar absorber (Fig. 1B).

To achieve the adjustable infrared spectroscopy of the SSAC device, multi-layer deposition was conducted and the thickness of each layer was shown in Fig. 2A. After depositing 200 nm $VO_2$ film on a 2-inch-diameter/500-μm-thick $Al_2O_3$ wafer with excellent infrared property (fig. S2), a 50-nm $Al_2O_3$ layer was covered on the top surface. In addition, a 200-nm metallic Al film was finally deposited on the back-side of the $Al_2O_3$ wafer, as shown in Fig. 2B and fig. S3. The SEM cross-section images for the device were shown in Fig. 2, C and D, which clearly showed the thickness of each layer. In the SSAC device, the 200 nm $VO_2$ film is important to tune the spectrum property due to its phase transition behavior. Temperature dependent Raman spectra in Fig. 2E was used to monitor the phase transition of $VO_2$ layer. The variations of sharp Raman peaks during the heating and cooling process clearly showing the reversible MIT behavior. Well uniform and epitaxial $VO_2$ thin films with high quality was also evidenced by the resistance distribution (fig. S4), X-ray diffraction (XRD) characterization (fig. S5), and the X-ray absorption near-edge spectroscopy (XANES) characterization (fig. S6). The resistance measurement as the function of temperature was shown in Fig. 2F and clear hysteresis behavior was observed. The sharp resistance change up to four orders of magnitude, further indicating the high quality of the deposited $VO_2$ layer on sapphire wafer. The differential curves in the insert clearly pointed out the critical temperature of 67.9 ℃ and 63.7 ℃ for the heating and cooling loops respectively, which was quite consistent with previous report for pure $VO_2$ epitaxial film(*32*). In fact, it was possible to decrease the critical temperature $T_c$ to room temperature or even lower by external atomic doping (W, Mo atoms etc) into the $VO_2$ layer(*31*), which should be used to adjust the working temperature of the proposed SSAC device.

According to the multi-layer structure of the SSAC in Fig. 2A, we simulated the absorptivity spectra before and after the phase transition in Fig. 3A. The results showed that when the temperature was lower than $T_c$, the device had very high emission in the atmospheric window, which was suitable for NRC mode to harvest the coldness from the out space. While the $T>T_c$, $VO_2$ layer converted to metallic rutile phase. From the spectra simulation, it was clear that the SSAC showed high absorption in the solar spectrum range (0.3−2.5 μm) and low emissivity in the mid-infrared band (2.5−25 μm). The simulation results indicated the SSAC could be realized based on the multi-layer structure in Fig. 2A.

The experiential measurements (fig. S7) also confirmed this SSAC structure design as shown in Fig. 3B−3E. The temperature dependent testing results showed that the absorptivity of the fabricated SSAC device had little variation in the solar spectrum (0.3−2.5 μm), while distinct change was observed within the atmospheric window (8−13 μm), which switched from high emission to low emission across the phase transition temperature of $T_c$. This pronounced optical change within the atmospheric window will exactly satisfy the SSAC device for continuous energy harvesting. During nighttime, the temperature of the SSAC is lower than $T_c$, the device has average emissivity of ~0.83 and ~0.75 in the solar radiation and atmospheric window region, respectively. However, if going to daytime, the temperature of SSAC rises after absorbing solar radiation. When the device temperature is higher than $T_c$, the phase transition of $VO_2$ layer will occur. Resultantly, the SSAC device switches off radiative cooling and becomes a solar absorber, which has an averaged absorptivity ~0.89 in the solar radiation band and ~ 0.25 in MIR band. This switching behavior only lies on the device temperature and no any extra energy input is required, exactly overcoming the intrinsic spectral confliction.

To explore the SSAC's performance under realistic weather conditions, we conducted an outdoor experiment (fig. S8−S10) on a clear summer day in Urumqi, China (43°50'21", 87°35'52"), by exposing it to the sky on a building roof throughout a 24-h day–night cycle and comparing its stagnation temperature to the ambient air temperature (Fig. 4). A typical measurement (Fig. 4A) shows the temperature of the SSAC, the vacuum chamber, the ambient air, and the solar irradiance. From the curves we can firstly observed that after the SSAC is exposed to the environment (shortly after 18:20 Beijing time), its temperature decreases rapidly and reaches equilibrium temperature. During the nighttime, the temperature of the SSAC is approximately −12 ℃, which is approximately 20 ℃ lower than the ambient air temperature (Fig. 4B), indicating the SSAC can achieve high-performance NRC. While it is observed that the SSAC's temperature increases rapidly after sunrise (around 10:00 am) due to its high absorptivity to sunlight. As shown in the inset of Fig. 4A, there is a sudden change of the temperature rising rate for the SSAC, while the rising rate of solar radiation is unchanged. The turning point corresponds to the phase transition of $VO_2$ film close to the critical temperature. The SSAC has larger temperature rising rate after phase transition, which means it shows larger DPT efficiency. In addition, the peak solar irradiation is ~800 W·m$^{-2}$ during the whole day, the SSAC's stagnation temperature is ~185 ℃ (~170 ℃ above the ambient temperature). The stagnation temperature of the SSAC demonstrates its efficient DPT conversion ability, below which it can output useful heat. At sunset, the DPT process degraded, resulting in the obvious decrease of device temperature. If the SSAC's temperature is lower than $T_c$, the SSAC device starts NRC mode spontaneously for coldness energy harvesting.

In addition, we further develop a theoretical model (fig. S11−S12) for the SSAC device to predict its thermal performance. Fig. 4, B and C compare the experimental results with our theoretical predictions. The model can be used to predict the steady-state temperature of the SSAC, and as compared against the observed performance of the SSAC (Fig. 4B). In the theoretical prediction, we bound the performance of the SSAC under maximal and minimal parasitic heat loss. The measured temperature of the SSAC fluctuates with time, which may be related to the wind and invisible cloud coverage. It is observed that the experiment agrees well with the theoretical simulation for the DPT mode in Fig. 4C. The theoretically predicted temperatures is mainly determined by the spectral characteristics of the SSAC before and after MIT transition.

Two clear turning points around the critical temperature of the $VO_2$ layer are reflected by the temperature change of SSAC device, further confirming the fact that the switching between DPT and NRC modes is mainly originated from the phase transition of $VO_2$ layer. Accordingly, if change the $T_c$ value of $VO_2$ layer (atomic doping, etc.), it is able to modulate the SSAC performance within a controllable temperature range, showing much more potentials for energy applications in different situations (winter/summer or cold zone/warm zone, etc.).

In a summary, we have achieved a spectral-self-adaptive coating (SSAC) based on $VO_2$ phase transition material, which principally overcome the infrared conflict for harvesting energy from the sun and outer space efficiently by the daytime-photothermal and nighttime-radiative-cooling routes. Our results demonstrate that the prepared SSAC device exhibits superior DPT and NRC ability, showing considerable potential for future energy applications.

**Acknowledgments:** We thank Dr. Mikhail A. Kats, Jie Tian and Chenghao Wan for their help on this study. This work was partially carried out at the USTC Center for Micro and Nanoscale Research and Fabrication. The approved beamtime on the XMCD beamline (BL12B) and Infrared spectroscopy/microspectroscopy beamline (BL01B) in National Synchrotron Radiation Laboratory (NSRL) of Hefei were also appreciated. **Funding:** This research was sponsored by the National Science Foundation of China (11574279, 5171101721, and 51776193), the funding supported by the Youth Innovation Promotion Association CAS and the Fundamental Research Funds for the Central Universities (108-4115100092). **Author contributions:** X.A. and B.L. contributed equally to this work. C. Z. and G.P. conceived the concepts and designed the experiments. B.L., H.R., Y.Y., Q.Z., L.L., and C.Z. performed the material preparation, characterization, and analysis. X.A., M.H., B.Z., and J.L. went to Urumqi for field test. X.A., H.Y., J.C., J.F., and G.P. developed the thermal performance model. X.A., B.L., C.Z., and G.P. wrote the manuscript. **Competing interests:** The authors declare no competing interests. **Data and materials availability:** All data is available in the main text or the supplementary materials. Information requests should be directed to the corresponding authors.

**Supplementary Materials:**

Materials and Methods

Supplementary Text

Figures S1-S12

References (*33-38*)

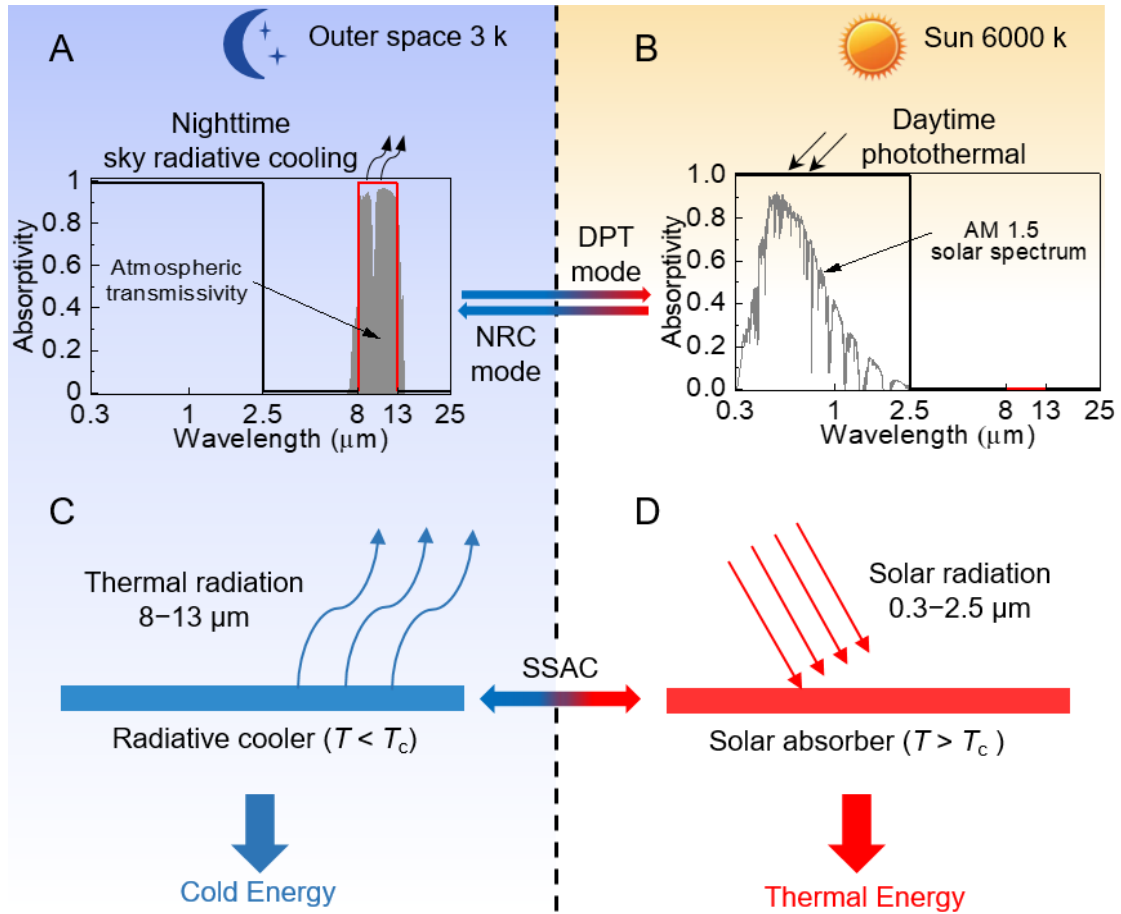

**Fig. 1. Principle of the spectral-adjustable NRC-DPT device.** Spectral absorptivity of spectral- adjustable NRC-DPT device in nighttime mode (**A**) and daytime mode (**B**). An ideal spectral-adjustable NRC-DPT emitter should be ideal photothermal collector in daytime and ideal radiative cooler in nighttime; (**C**, **D**) Schematic diagram of the SSAC. $T$ is the temperature of the SSAC, $T_c$ is the critical temperature of the $VO_2$ layer. When $T < T_c$, the SSAC is a radiative cooler. It will change to a solar absorber when $T > T_c$.

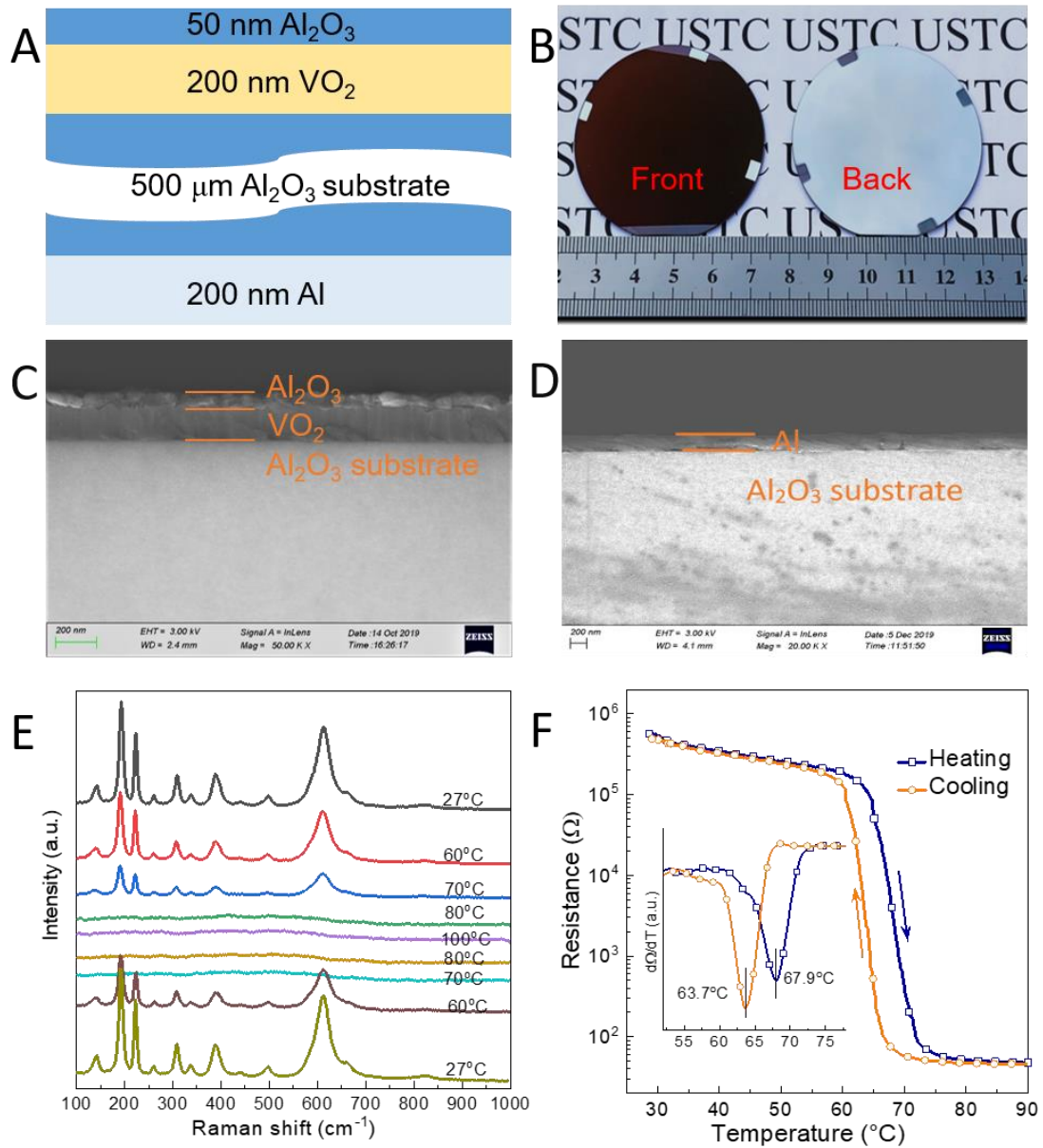

**Fig. 2. Material characterization of the SSAC.** (**A**) Multi-layer deposition and the thickness of each layer for the SSAC device. (**B**) The optical images for the SSAC device. (**C**, **D**) The SEM cross-section images for the multi-layer depositions on 2-inch-diameter and 500-μm-thick $Al_2O_3$ wafer. The 50 nm $Al_2O_3$/200 nm $VO_2$ layers were deposited on the front side while 200 nm metallic Al layer was deposited on the backside. (**E**) The Raman spectra for the 200 nm $VO_2$ layer on $Al_2O_3$ wafer at different temperature, showing the clear phase transition behavior at the critical temperature. (**F**) The resistance measurement as the function of temperature (R-T) during the heating and cooling process, showing the insulator-metal phase transition with clear hysteresis behavior. The insert shows the related differential curves, showing the critical temperature of 67.9 ℃ and 63.7 ℃ for the heating and cooling loops respectively.

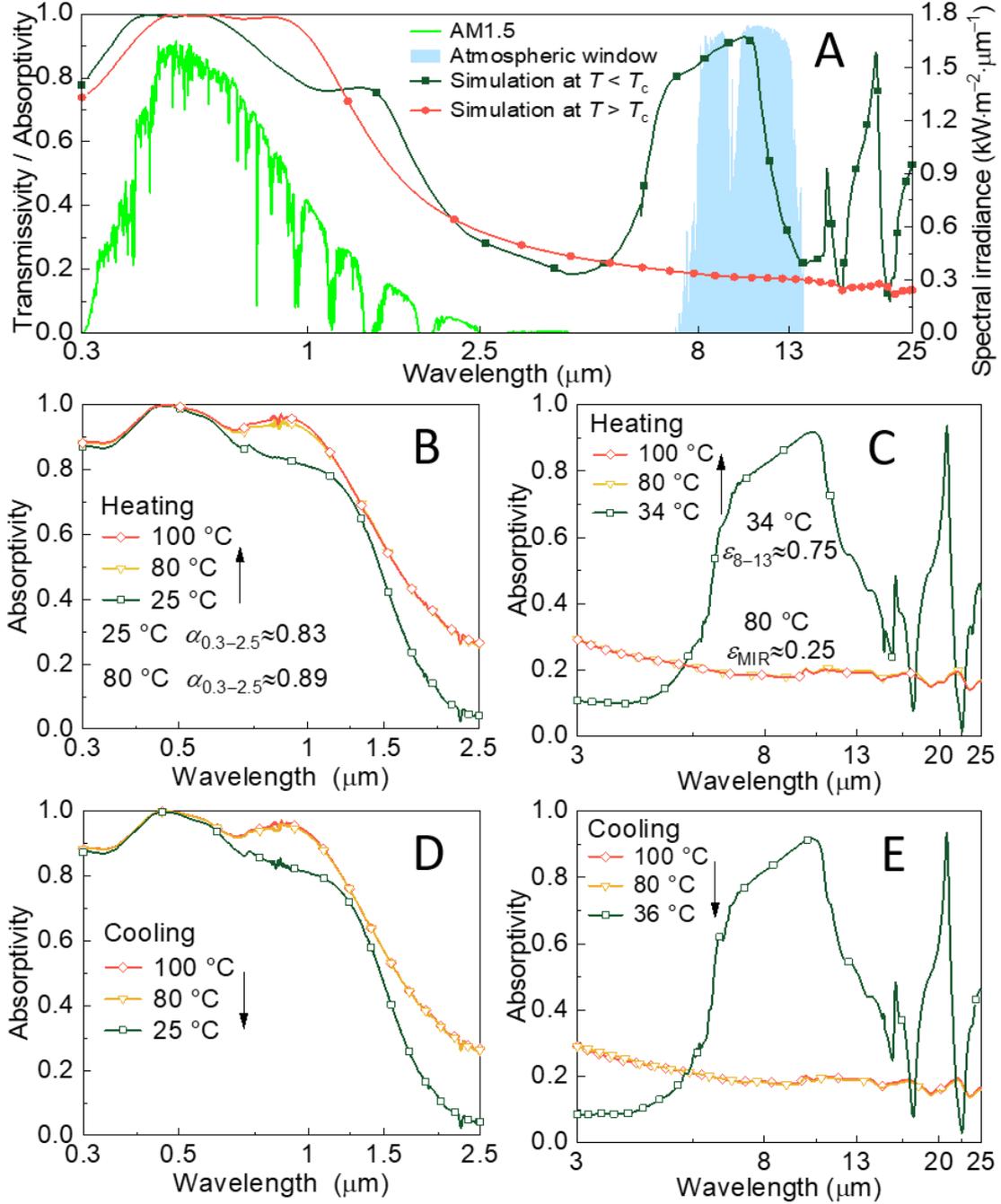

**Fig. 3. Optical characterization the SSAC.** (**A**) Spectra simulation based on the multi-layer structure of the SSAC device as shown in Figure 2A, showing the self-adaptive property of spectra in the atmospheric window (8−13 μm). (**B**, **C**) Experientially measured absorptivity of the SSAC in the solar and mid-infrared (MIR) band spectrum during the heating process. The absorptivity in the solar spectrum (0.3−2.5 μm) showed little change, while great change was observed within the atmospheric window (8−13 μm), showing the pronounced self-adaptive property. (**D**, **E**) showed the absorptivity of the SSAC device during the cooling process, showing the similar self-adaptive property within the atmospheric window.

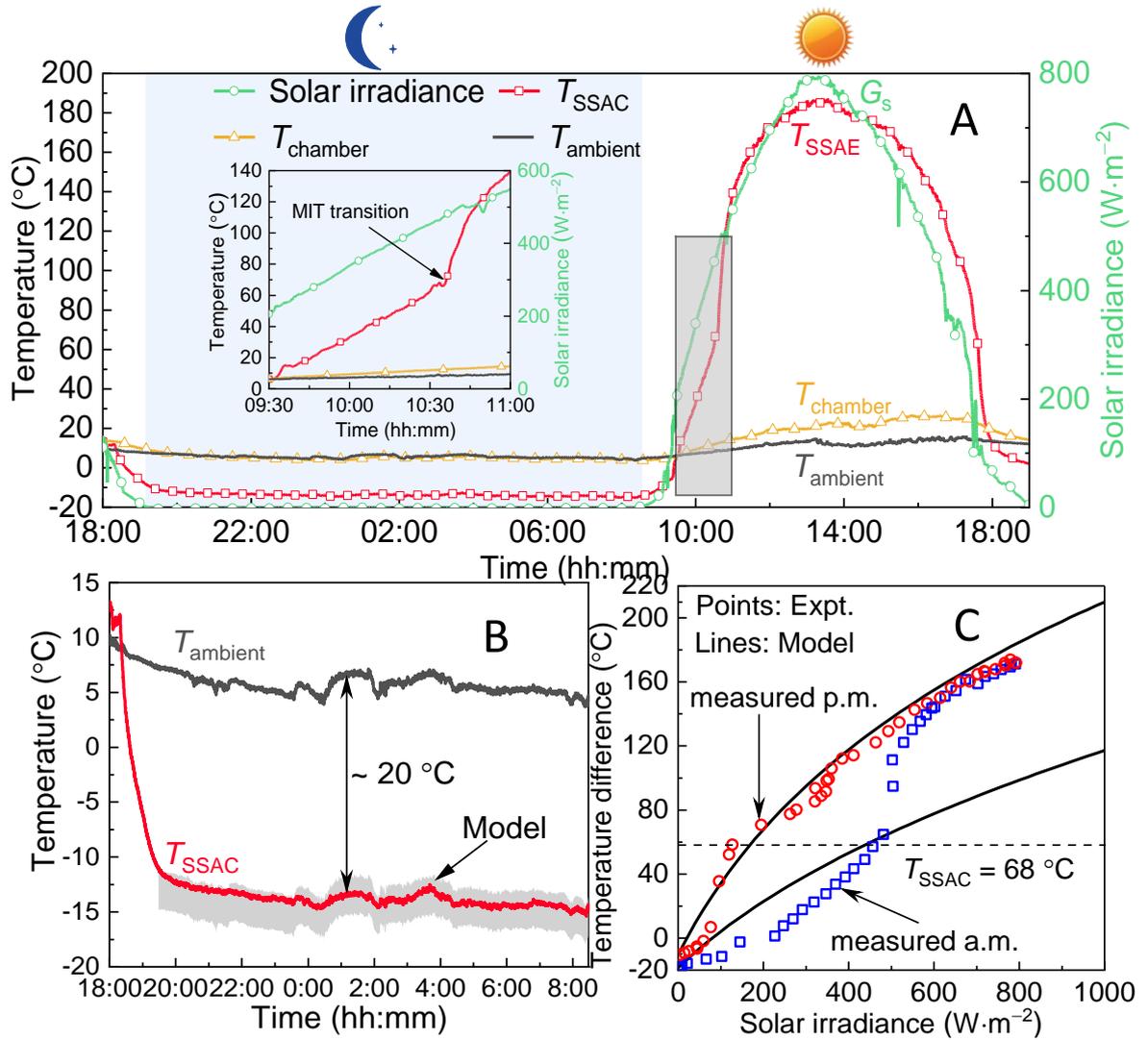

**Fig. 4. Nighttime sky radiative cooling and daytime photothermal performance of the SSAC.** (**A**) Twenty-four-hour continuous measurement of steady-state temperature of the emitter in Urumqi. Inset shows the MIT transition process of the SSAC. The temperature of the SSAC (red), the outer surface of vacuum chamber (yellow), the ambient air (grey), and the solar irradiance (blue, right axis) are presented. (**B**) Nighttime sky radiative cooling performance. Theoretical model is used to calculate the real-time stagnation temperature of the SSAC given observed air temperatures. The theoretical model bound the cooling performance of the SSAC under maximal and minimal parasitic heat loss. (**C**) Comparison experimental result with theoretical predictions under daytime conditions. The circled area represents the critical temperature zone. Two simulation line correspond to the spectral properties of the SSAC before and after MIT transition. The ambient air temperature is assumed to be 15 °C in this model.

15